\documentclass{article}

\usepackage[english]{babel}

\usepackage[letterpaper,top=2cm,bottom=2cm,left=3cm,right=3cm,marginparwidth=1.75cm]{geometry}

\usepackage{amsmath}
\usepackage{graphicx}
\usepackage[colorlinks=true, allcolors=blue]{hyperref}
\begin{document}

\title{Depth estimation of endoscopy using sim-to-real transfer}
\author{Bong Hyuk Jeong\thanks{Bong Hyuk Jeong is with the Department of Health Sciences and Technology, GAIHST Gachon University,Incheon, Korea (e-mail: dkalshzz12@gachon.ac.kr).}, 
Hang Keun Kim\thanks{Hang Keun Kim is with the Department of Biomedical Engineering, Gachon University,Incheon, Korea (e-mail: dsaint31@gachon.ac.kr).}
and Young Don Son \thanks{Young Don Son is with the Department of Biomedical Engineering, Gachon University,Incheon, Korea (e-mail: ydson@gachon.ac.kr).}
\thanks{Date of current
version december 26, 2021. This research was supported by a grant of the Korea Health Technology R\&D Project through the Korea Health Industry Development Institute (KHIDI), funded by the Ministry of Health \& Welfare, Republic of Korea (grant number: HI19C0656030021). (Corresponding authors:Young Don Son)}}
\maketitle
\begin{abstract}
In order to use the navigation system effectively, distance information sensors such as depth sensors are essential. Since depth sensors are difficult to use in endoscopy, many groups propose a method using convolutional neural networks. In this paper, the ground truth of the depth image and the endoscopy image is generated through endoscopy simulation using the colon model segmented by CT colonography. Photo-realistic simulation images can be created using a sim-to-real approach using cycleGAN for endoscopy images. By training the generated dataset, we propose a quantitative endoscopy depth estimation network. The proposed method represents a better-evaluated score than the existing unsupervised training-based results.
\end{abstract}

\section{Introduction}
In robotics, the simultaneous localization and mapping(SLAM)\cite{b1} technique has been developed to generate a spatial map from the data acquired while exploring unknown areas. This map provides important navigation information in driving. In general, SLAM technology collects the spatial position information measured from the odometer as well as the depth information measured from the stereo-camera or several distance sensors. SLAM technology is used for navigation and is a representative technology in the field of computer vision under the name of visual SLAM using a camera. This SLAM technology replaces GPS technology and is widely used in autonomous driving and drones.
Skillful operators can manipulate the endoscope depending on several recognizable anatomical landmarks for them in real-time camera images, and, thereby, this navigational map technology is not necessary in use of endoscopy for diagnosis or surgical operation,  However, it can help the operators recognize the location of the endoscope more easily while manipulating. Under the gastrointestinal surgery environment, the three dimensional (3D) navigation map enables the operators to mark the lesional location and shows to the operators if the surgery device is approaching near the marked lesion. Furthermore, in the case of wireless capsule endoscopy\cite{b2} being developed, unlike general endoscopy, the navigation map is essential to manipulate it in real-time. 
Generating 3D maps of the gastrointestinal region by the SLAM technology will surely improve the navigational function in operating the endoscope. In order to implement this, the depth or distance information of the corresponding endoscopic images should be measured by the endoscopy. in hardware-wise, such as stereo cameras or time-of-flight(TOF) cameras. However, it increases the size of the camera module of the endoscope and, thereby, it is currently challenging issues for developing the SLAM technology in the endoscopy. 
Therefore, several deep learning-based approaches, such as convolutional neural networks (CNN), were recently proposed in order to estimate the depth information from the images instead of hardware-wise approaches. Unsupervised learning\cite{b3} was used for estimating the depth from the RGB image and camera movement data, which showed the correlation to depth. But, It was impossible to evaluate the accuracy of the result, because of  non-existence of ground truth. Other groups generated both endoscopy and depth images through simulation\cite{b4}. However, the simulated endoscopy images were not realistic enough to train the deep learning and, thereby, the performance was not satisfactory when it was applied to the real endoscopy images. There is a sim-to-real transfer method as an effective method when applied in real-world situations in simulations. A method that is mainly used in tasks that take a lot of time to collect data, such as robot motion planning using reinforcement learning\cite{b5}, is applied to real-world situations using deep learning that translates the simulated environment into a real-world environment. As discussed above, generating the realistic endoscopy images as well as the corresponding depth images from the simulation environment is the most important for training the depth estimation network from the real endoscopy images In this work, we propose a novel method generating a pair of 2D models of colonoscopy and depth images, which were generated from the simulated 3D model from the CT data, and translating the model image to the realistic colonoscopy data by using the cycleGAN. A pair of realistic colonoscopy images and the corresponding depth images were then used to train the deep learning network which predicts the depth information from the real colonoscopy images. In addition the prediction errors were also quantitatively evaluated using the simulated test datasets.

\section{Methods}
\subsection{Overall processing procedures}
The entire process of the training and prediction process for the depth estimation of the endoscopy was shown in Figure 1. We generated a colon 3D model  from CT Colonography via segmentation process. Colon models were constructed in the Unity environment for endoscopy simulation. Endoscopy simulation produces the ground truth of computed depth images as well as endoscopy images of RGB. The generated data is trained with cycleGAN to map textures such as real endoscopy, performing a sim-to-real process. The final translated data and depth ground truth images are learned with the residual Unet, allowing depth estimate in real-world endoscopy images.

 \begin{figure*} [!th]
\begin{center}
\includegraphics[width=0.98\linewidth]{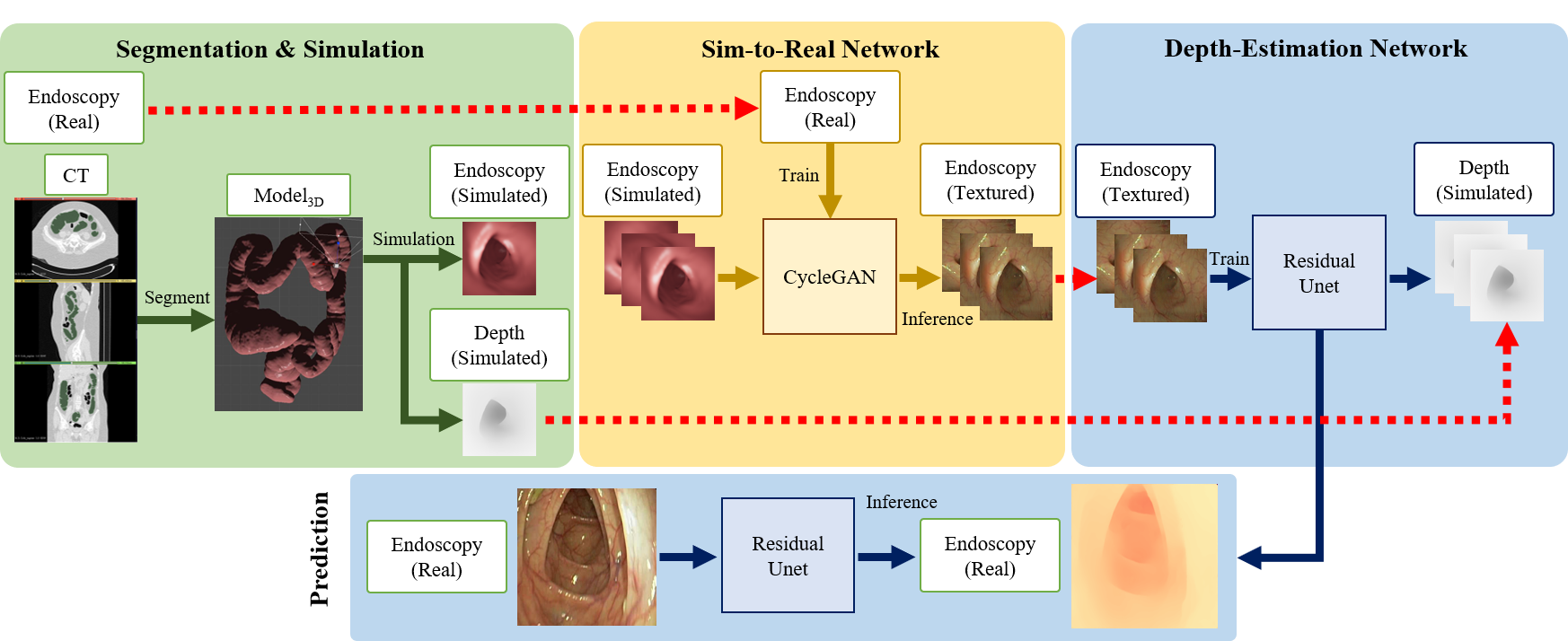}
\end{center}
\caption{Overall processing diagram}
\label{fig:short}
\end{figure*} 

\subsection{System environment}
The system for the deep learning training had Intel Core i7-5930K (3.5 GHz, 6 cores, 12 threads) CPU and DDR4 2133 (64 GB = 8 x 8)memory, and NVIDIA GeForce GTX TITAN X 12GB GPU. Tensorflow 1.13 and Keras 2.2.4 were used as the deep learning framework.

\subsection{Generation simulation endoscopy Images}
\textbf{CT colonography dataset} To extract the colon model, a CT colonography dataset, a public dataset from the Cancer imaging archive site\cite{b6}, was used. Among 836 CT scans, eight scans were randomly selected. The dataset of CT images included both supine  and prone poses, but supine images were only used in this study, because the colon in the prone posture was pressed by the human weight of the organ and thereby it was difficult to segment the colon. All images were oriented in axial direction and the number of slices per subject was in a range from 523 to 628 slices. The voxel size of the image was 0.78 x 0.78 x 1 mm3, and the image size was 512  x 512 pixels. The CT scanner was Somatom Sensation 16 (Siemens, USA). More detailed description of the dataset is found in Smith K et al.\cite{b7} and Johnson, C et al.\cite{b8}. 

\textbf{Extraction of the three-dimensional colon model from CT} To generate the colonoscopy images as well as the corresponding depth images, a 3-D colon model was reconstructed from the CT images. Using the Slicer4\cite{b9}\cite{b10}, the colon was segmented in a semi-automatic way. The thresholding parameters used in the segment editor are set from the minimum value to 1/3 in the CT values range. Fine adjustment for the manual segmentation was required for each data with visual inspection. Among the segmented areas, the area having less than 100,000 voxels were excluded because of the anatomical continuity of the colon. To generate surface data of the colon, the segmented model was expanded three voxels toward the outside from the surface and, then, the segmented model was subtracted from the expanded model. The model was saved in 3D mesh data.

\textbf{Generating simulation endoscopy dataset} The Unity software\cite{b12} provided the simulation tool for the colonoscopic environment from the segmented CT image. The extracted colon models of mesh form were loaded into the Unity and driven by controlling the camera of the UI user interface in the Unity to extract a pair of endoscopic view and depth images. The endoscopic image was captured as an RGB image as a still scene of endoscopic view and the corresponding depth images were automatically generated by the Unity software. Approximately 2000 RGB and depth images were extracted from each colon model and, thereby, a total of 16,369 pairs of images were obtained.
 In the simulation configurations, the camera parameters were set to have 256-by-256 pixels and 120 degrees of field of view (FOV). Several camera effects in the endoscopic environment were added, such as motion blurring by camera movement, depth of field by focal length, and light reflection. Simulated endoscopy images were generated along with the colon model with a specific position and angle of the camera operated by using a keyboard and mouse. Depth images are extracted at the same location as the RGB camera, and a value of 0.01 to 20cm is converted to 8 bits and stored. This simulation process started from the end of the colon and continued until arriving at the other end and returned to the starting point. It is in the form of a video, but it is stored by frame, at 30 frames per second.

\begin{figure}[!h]
\includegraphics[width=\linewidth]{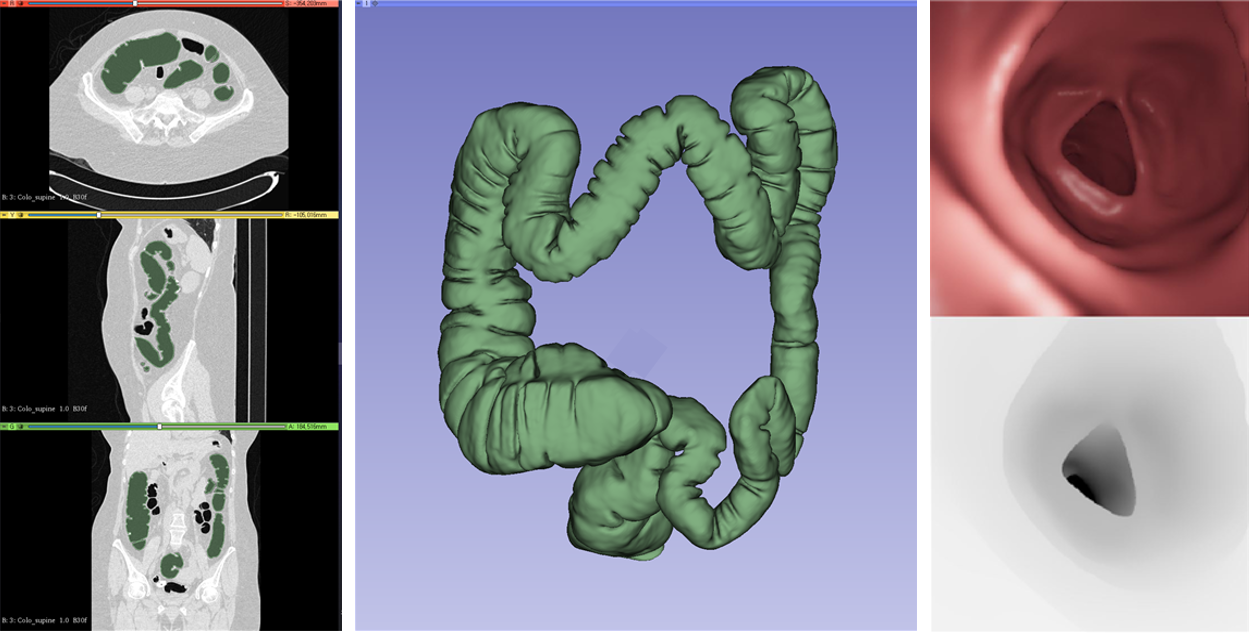}
\caption{Segmentation processing in colonography CT, segmented colon model, endoscopic camera, and depth camera in simulation}
\label{fig:onecol}
\end{figure}

\subsection{Translation from simulation domain to real domain}
 The simulation endoscopy images extracted through simulation had no texture of a real colon image. To solve this problem, we map the textures of real endoscopy to simulation data using a domain translation network. Among domain translation networks, we used CycleGAN (Cycle-Consistent Adversarial Networks)\cite{b13} specialized for unpaired data translation. CycleGAN enables bi-directional translation, but we only use the simulation to real generator for inference.

\textbf{Sim-to-real networks (CycleGAN)} CycleGAN consists of 4 convolutional networks: generator $G$, generator $F$, discriminator $D_x$, and discriminator $D_y$. The Sim-to-Real generator, $G$, translates the simulation image domain, $X$, into the real image domain, $Y$, and the Real-to-Sim generator, $F$, translates the domain $Y$ to $X$. The domain data of $X$ and $Y$ are defined as $\left\{x_{i}\right\}_{i=1}^{N} \in X$ and $\left\{y_{i}\right\}_{i=1}^{M} \in Y$, and the samples are $x$ and $y$.The two adversarial discriminators $D_x$ and$D_y$ discriminate the image of each domain as real domain images, and the image translated by $G$ and $F$ as fake domain images. 

\begin{figure}[t]
\includegraphics[width=\linewidth]{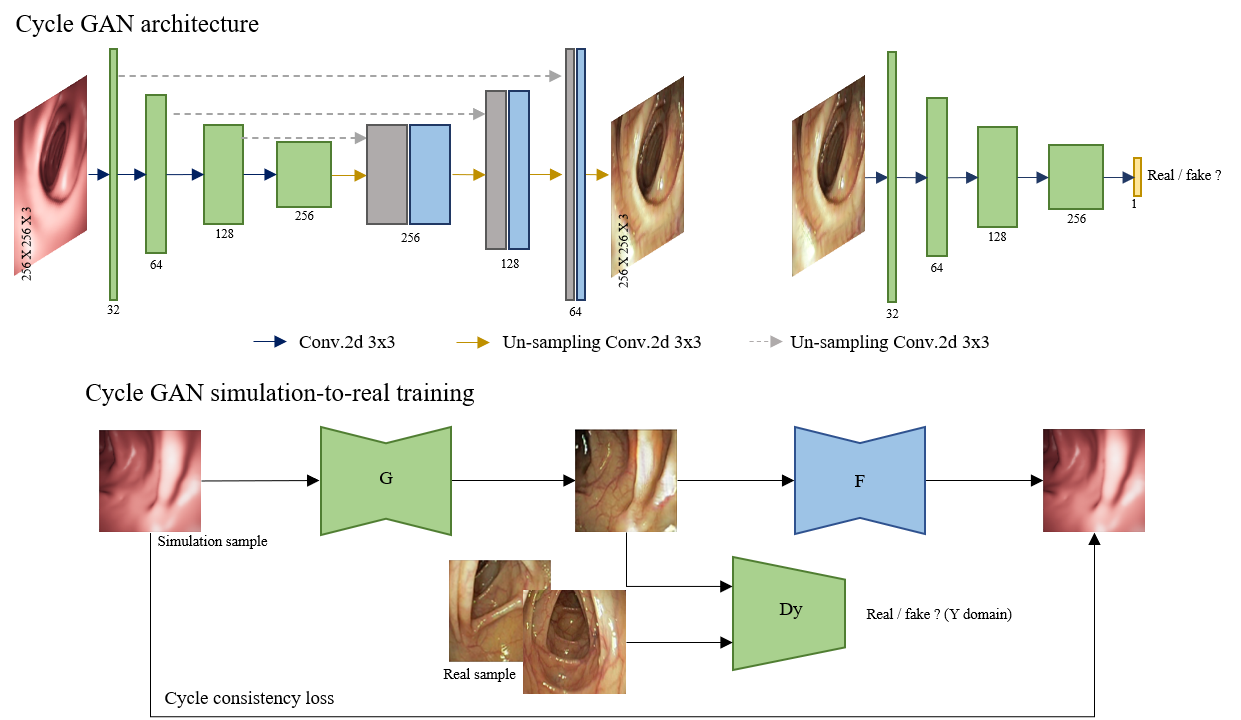}
\caption{CycleGan generator and discriminator architecture and training phase}
\label{fig:long}
\label{fig:onecol}
\end{figure}

Adversarial loss function includes the translation in both directions. For example simulation-to-real loss is:
\begin{equation}
\begin{aligned}
\mathcal{L}_{G A N}\left(G, D_{Y}, X, Y\right)= E_{y \sim Y}\left[\log D_{Y}(y)\right] + E_{x \sim X}\left[\log \left(1-D_{Y}(G(x))\right)\right]
\end{aligned}
\label{eq}
\end{equation}
The simulation-to-real generator G is minimized in the loss function to generate a realistic Y domain image by Dy, and Dy is maximized. CycleGAN includes cycle consistency loss. This represents $x \rightarrow G(x) \rightarrow F(G(x)) = \hat{x}$ and $y \rightarrow F(y) \rightarrow G(F(y)) = \hat{y}$ in which the each initial domain input passes through two generators and returns to the original domain image.
\begin{equation}
\begin{aligned}
\mathcal{L}_{cycle}(G, F)= E_{x \sim X} \left \|(F(G(x)), x)\right \|_{2} + E_{y \sim Y}\left \|(G(F(y)), y)\right \|_{2} 
\end{aligned}
\label{eq}
\end{equation}
The final loss is to adjust the GAN loss and cycle loss in weight is represented by the following formula:
\begin{equation} 
\mathcal{L}_{cycleGAN} = \alpha \mathcal{L}_{G A N}\left(G, D_{Y}, X, Y\right) + \beta \mathcal{L}_{\text {cycle }}(G, F)
\end{equation}
$\alpha$ and $\beta$ can adjust the weights of the existing adversarial loss and  the cycle consistency loss, respectively, as described in Jun-Yan Zhu et al\cite{b13}. We used  $\alpha$ and $\beta$ as 1 and 10, respectively, as optimized parameters.
Each of the two generators and discriminators has the same network structure. The generators use the Unet\cite{b17} structure. The configuration of the layers of the encoder is convolution with 4-by-4 kernels, a leaky Relu activation function that allows 0.2 negative numbers, and an instance normalization order. The filter increases by 32 for each convolution, and the encoder has 4 layers. The decoder layer has the same configuration as the encoder, but there is a bilinear interpolation that doubles before input. The decoder also contains a concatenation path that connects to the encoder output features of the same size. After the convolution, the final output passes through the tanh activation function and is output in the same size as the input image. The discriminator has the same configuration layer as the generator, but the filter expands by 64. The final output uses the patch discriminator used by  P. Isola et al\cite{b21}.

\textbf{Data discrimination} We recognized that some of the generated data were different from real endoscopu. The reason is that the real endoscopy does not move freely because there is a line, the shooting range is limited. However, in the simulation, the movement is free, so the white image is sampled because it is too close to the wall, or the dark image  is sampled because it is too far from the wall. To discriminate this, the color of the image was converted into HSV format and the data of the top 10\% and the bottom 10\% were deleted based on the brightness. The number of deleted data is 14,914 by removing 2,258 out of a total of 17,172.

\textbf{Data preprocessing for training} Two data sets were used for cycleGAN training. For the real image dataset, a single subject was selected from the transverse colon video dataset\cite{b14}, and 1,034 images were captured from the ascending and transverse colon video. The simulation data set used data extracted from one person's colon model, and there are a total of 2,116 images. The video had 640 x 480 pixels and 25 frames per second and they were converted to image frames and down-sampled to 256 x 256 images. All images used were normalized in a range from -1 to 1, and horizontal, vertical flipping, and color space random shifting was used for data augmentation.

\textbf{Translating simulation data} CycleGAN has two generators that can be bi-directional, but we only use the generator of simulation to real. We translate simulation images of 14,464 except for 2,116 images used for CycleGAN training, which we use for depth estimation network training and testing.

\textbf{Networks experiments} CycleGAN's optimizer used Adam with a running rate of 0.0002, beta1 of 0.5 and beta2 of 0.99. The batch size was 16 and epochs were trained up to 300, but the generator loss minimum of 220 epoch was taken and used.

\subsection{Depth estimation using residual Unet}
The task of estimating the depth is obtained by calculating the distance as stereo input in computer vision. Nikolaus Mayer et al.\cite{b15}, who implemented this through deep learning and showed meaningful results, uses the stereo input. Therefore, there are many studies on estimating depth from monocular images\cite{b16}. In this study, since depth estimation networks are being studied based on Unet, depth estimation was performed using residual Unet\cite{b18} with a residual structure.

\textbf{Residual Unet Network} As the network architecture of the depth-estimation, a residual Unet\cite{b18} was used. The residual Unet is the most frequently used in the segmentation of medical images. The Residual Unet architecture maintains the concatenate path of the existing Unet\cite{b17}, and replaces all convolution layers with residual cnn blocks. In order to propagate the depth of the feature, the identity path is expanded by 1-by-1 convolution, and the residual path is expanded only by the last of the two convolutions of the 3-by-3 kernels. The filter extends 64 per layer. The final output passes through the convolution of the 1-by-1 kernel and the activation function of the sigmoid and has the same size as the input image of a channel. 

\begin{figure}[htb!]
\includegraphics[width=\linewidth]{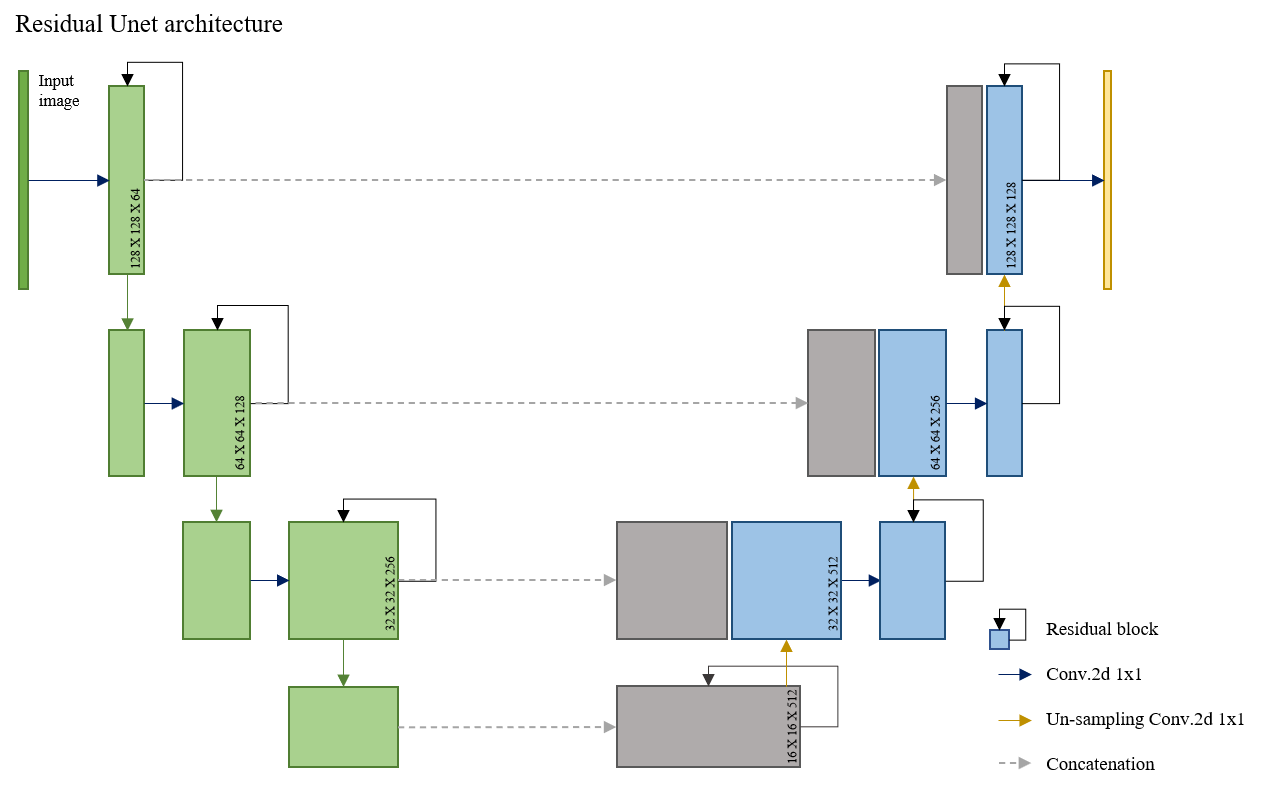}
\caption{Residual Unet architecture}
\label{fig:long}
\label{fig:onecol}
\end{figure}

Generally, MSE(mean square error) is used as a loss function to measure errors in images such as depth, but we added SSIM(structural similarity index) to emphasize structural edges.

\begin{equation}\begin{aligned}\operatorname{MSE}=1 / n \sum_{i=1}^{n}\left(y_{i}-\hat{f}\left(x_{i}\right)\right)^{2} \end{aligned}\label{eq}\end{equation}
\begin{equation}\begin{aligned}\operatorname{SSIM}(x, y)=\frac{\left(2 \mu_{x} \mu_{y}+C_{1}\right)\left(2 \sigma_{x y}+C_{2}\right)}{\left(\mu_{x}^{2}+\mu_{y}^{2}+C_{1}\right)\left(\sigma_{x}^{2}+\sigma_{y}^{2}+C_{2}\right)} \end{aligned}\label{eq}\end{equation}
\begin{equation}\begin{aligned}\operatorname{Loss} = \alpha Loss_{MSE} + \beta Loss_{SSIM} \end{aligned}\label{eq}\end{equation}

The weights of MSE and SSIM are alpha and beta, respectively, and grid search was used to find the optimal weight parameters. MSE was fixed and SSIM was changed as [0.001, 0.01, 0.1, 0, 1, 10, 100].

\textbf{Data preprocessing for training} The generated endoscope images and depth images by the sim-to-real generator G were used as input and output images in the depth-estimation network with simulation ground truth depth images. Among the 14,914 translated data, 10,553 images were used for the training dataset, 2,116 images for the validation dataset, and 2,245 images for the test dataset. All images used were normalized pixels from 0 to 1, and horizontal, vertical flipping, color space random shifting was used with the augmentation technique in training.  

\textbf{Networks experiments} Residual Unet's optimizer for estimation depth images also used Adam with, running rate of was 0.0002, beta1 of 0.9, and beta2 of  0.99. Batch sizes were 16 and epochs were trained up to 300, but were used with weights of 150 epoch, the point where validation loss increased.

\section{Results}
\subsection{Data generation and translation}
\textbf{Translation from simulation domain to real domain} CycleGAN was used to map the texture of the colonoscopy images to the simulation endoscope images. A score is needed to compare the translated endoscope images with the real endoscope images. Recently, the GAN paper uses the Fréchet Inception Distance\cite{b19} (FID) evaluation score to quantitatively represent performance scores. FID measures performance using the InceptionV3\cite{b20} network. Features are extracted from real and translated data through the InceptionV3 network, and the distance between the data is calculated using the mean and covariance. 
\begin{equation}\begin{aligned} \mathrm{d}(\mathrm{X_{feature}},\mathrm{Y_{feature}})=\left(\mu_{\mathrm{X}}-\mu_{\mathrm{Y}}\right)^{2}+\left(\sigma_{\mathrm{X}}-\sigma_{\mathrm{Y}}\right)^{2}\end{aligned}\label{eq}\end{equation}

FID compared the real endoscopy dataset and the simulation dataset, and the real endoscopy data set and the translated dataset. These are indicated as simulation data and translated data in table 1. Since FID means that the smaller the value, the more similar, the translated dataset can be said to be closer to the real dataset.
\begin{table}[htb!]
\centering
\begin{tabular}{|l|l|}
\hline
\multicolumn{1}{|c|}{\textbf{Dataset}} & \multicolumn{1}{c|}{\textbf{FID}} \\ \hline
simulation data                        & 15.032                            \\ \hline
translated data                        & \textbf{4.663}                            \\ \hline
\end{tabular}
\caption{Scores of compared FID}
\label{tab:my-table}
\end{table}

\begin{figure}[htb!]
\includegraphics[width=\linewidth]{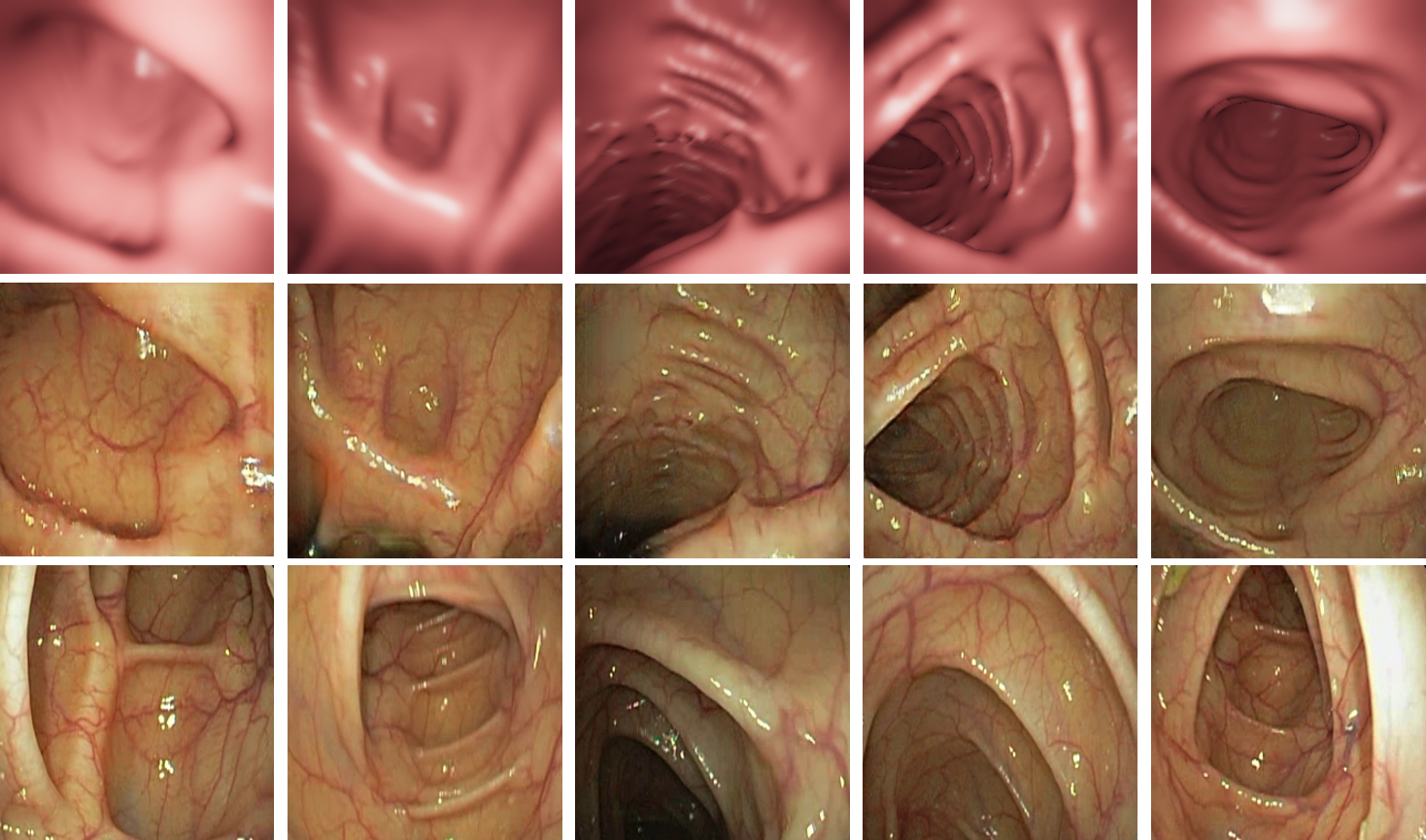}
\caption{The top is Simulation image sample of CycleGAN training images and the middle is a translated endoscopy image sample of CycleGAN result images and the bottom is an image sample of CycleGAN training images.}
\label{fig:long}
\label{fig:onecol}
\end{figure}

In Fig. 5, the top is a sample of the simulation image, and the middle is a translation of the top simulation image into a cycleGAN generator. The bottom image is a sample of real endoscopy images that are highly similar to the translated simulation images. The image is part of 14,464 data not used for cycleGAN training. The image shows that textures such as light reflection, blood vessels, etc. are synthesised.  

\textbf{Depth estimation} We evaluated the trained residual Unet a total of three models using 2,245 images of test data. The evaluated metrics were absolute relative error, squared relative error, root mean squared error, log scale root mean squared error and percentage error using in C.Godard et al.\cite{b16}. The results compared the ground truth of the test data with the depth prediction results of the comparative models. The two models are the same as the residual Unet, but the data used for learning are different. The data used for learning are simulation data and simulation data translated into cycleGAN. Lastly SC-SfMlearner \cite{b11} is an unsupervised learning-based method. A method called geometry consistency loss is a transformation matrix for camera movement that computes depth information after movement by applying it to depth information before movement. The data used for learning used translated simulation data, which can be learned in SC-SfMearner because it is in video format. For learning, the camera intrinsic matrix used the camera intrinsic matrix set by unity. The principal point  is 128, the principal point is 50 and the principal point is 1. Table.2 shows metrics comparing two comparative models with the proposed model. The bold type displays the best performance of metrics.

\begin{table*}[]
\centering
\begin{tabular}{|l|l|l|l|l|l|l|l|}
\hline
\textbf{}                                                                                    & \multicolumn{1}{c|}{\textbf{abs\_rel}} & \multicolumn{1}{c|}{\textbf{sq\_rel}} & \multicolumn{1}{c|}{\textbf{rmse}} & \multicolumn{1}{c|}{\textbf{rmse\_log}} & \multicolumn{1}{c|}{\textbf{a1}} & \multicolumn{1}{c|}{\textbf{a2}} & \multicolumn{1}{c|}{\textbf{a3}} \\ \hline
SC-sfm learner                                                                               & 0.900                                  & 0.648                                 & 9.801                              & 0.280                                   & 0.931                            & 0.989                            & 0.994                            \\ \hline
\begin{tabular}[c]{@{}l@{}}Residual Unet\\ training with simulation\end{tabular}             & 0.734                                  & 0.702                                 & 9.902                              & 1.098                                   & 0.756                            & 0.879                            & 0.905                            \\ \hline
\begin{tabular}[c]{@{}l@{}}Residual Unet\\ training with translation \end{tabular} & \textbf{0.384}                         & \textbf{0.302}                        & \textbf{6.361}                     & \textbf{0.161}                          & \textbf{0.991}                   & \textbf{0.996}                   & \textbf{0.998}                   \\ \hline
\end{tabular}
\caption{Scores of compared FID}
\label{tab:my-table}
\end{table*}

Figure.6 shows the input image and the inference result of the models, and the ground truth images. SC-SfMlearner is an unsupervised learning method that predicts depth images that lack detail compared to ground truth. Simulation images are vulnerable in environments such as blood vessels and light reflections commonly seen in endoscopy. 

\begin{figure}[htb!]
\includegraphics[width=\linewidth]{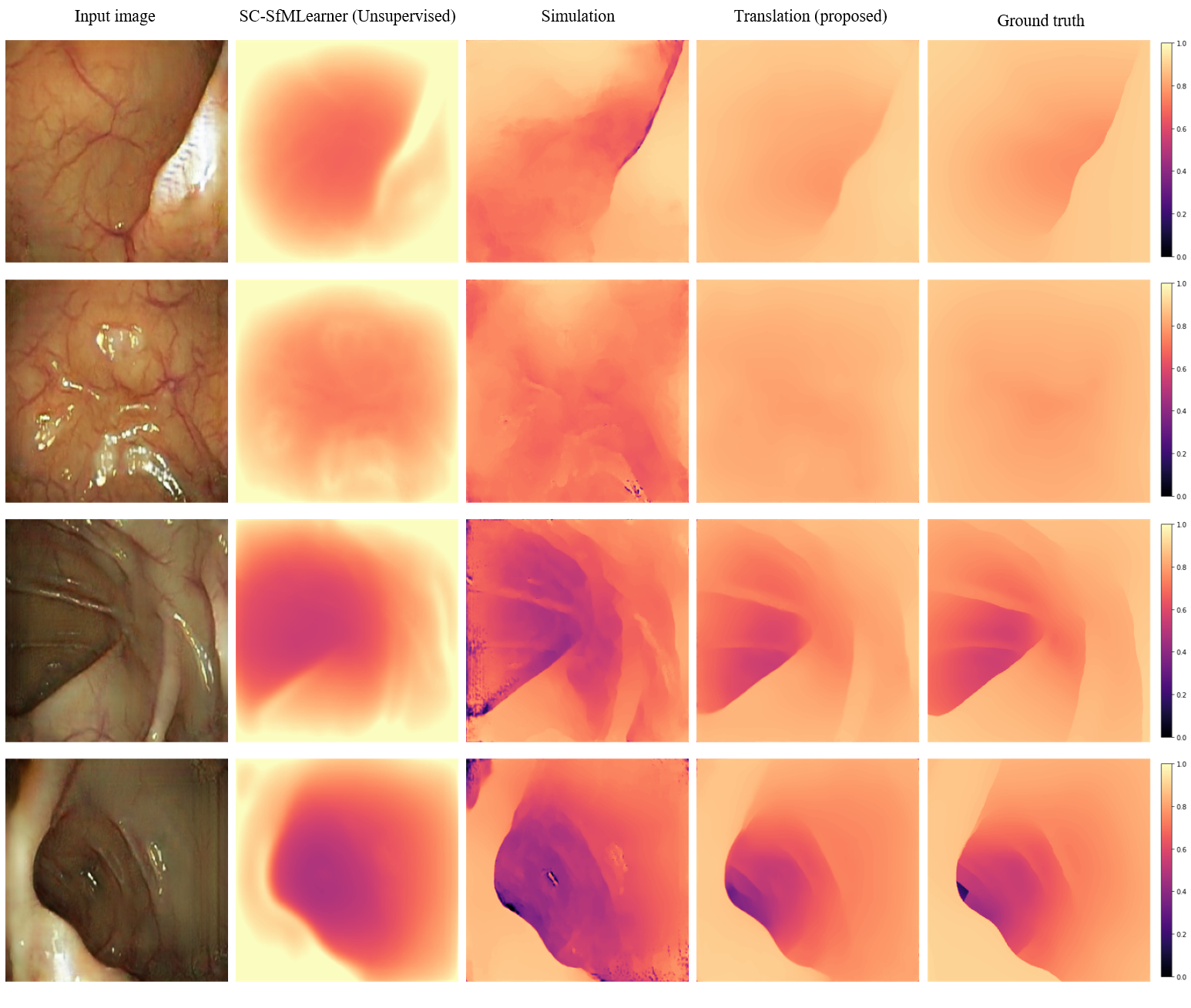}
\caption{Results of test data samples are shown. Input images in order from the left, SC-SfMLearner's inference, residual Unet's inference trained from simulated data, residual Unet's inference trained from translated data, and the last column is ground truth.}
\label{fig:long}
\label{fig:onecol}
\end{figure}

Figure.7 shows the results of the application to the real endoscopy and there is no ground truth. The problem with the resulting inference is that the real endoscopy has a circular letter box, but it was not considered at the time of use in the simulation environment, so the image of removing the letterbox was inputted. 

\begin{figure}[htb!]
\includegraphics[width=\linewidth]{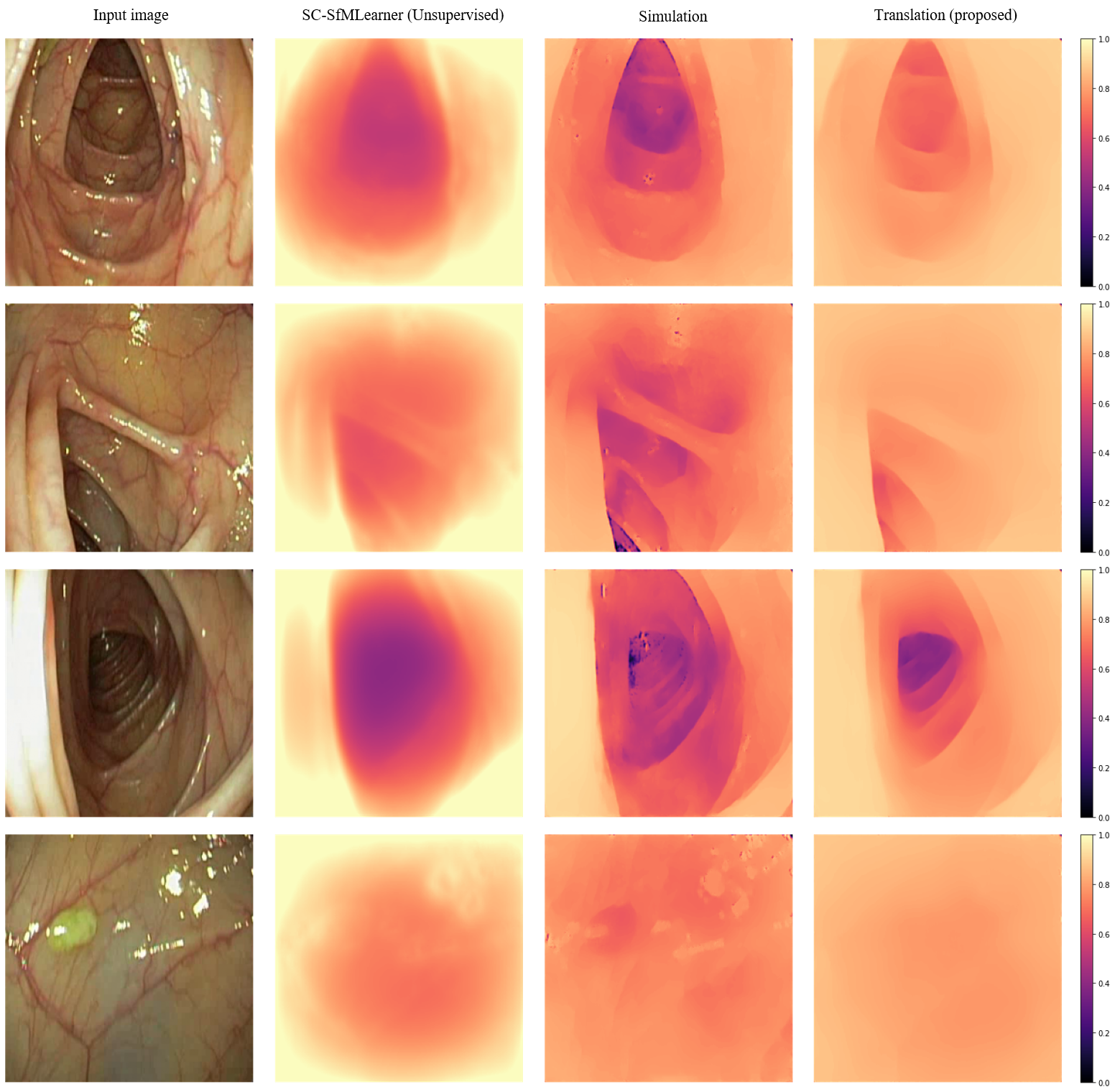}
\caption{ Results of test data samples are shown. Input images in order from the left, SC-SfMLearner's inference, residual Unet's inference trained from simulated data, residual Unet's inference trained from translated data, and the last column is ground truth.}
\label{fig:long}
\label{fig:onecol}
\end{figure}

\section{Discussion}
We performed analysis at the predicted depth image level to determine whether the depth estimation network was effectively trained through the translated dataset. In particular, we checked whether there was a depth error according to the texture, and whether there was a depth error when recognizing heuristics.

Figure.8 compares the inference results of the model trained with the simulation dataset and the model trained with the translated dataset. The left side of figure.8 shows the error of depth estimation for blood vessels. The simulation inference result is estimated to have appeared as a depth error by recognizing blood vessels as steps in the colon structure. However, the inference result of the model trained with the translated dataset was recognized as noise because blood vessels were well generated in the translated images, and the depth was estimated by ignoring the blood vessels.

\begin{figure}[htb!]
\includegraphics[width=\linewidth]{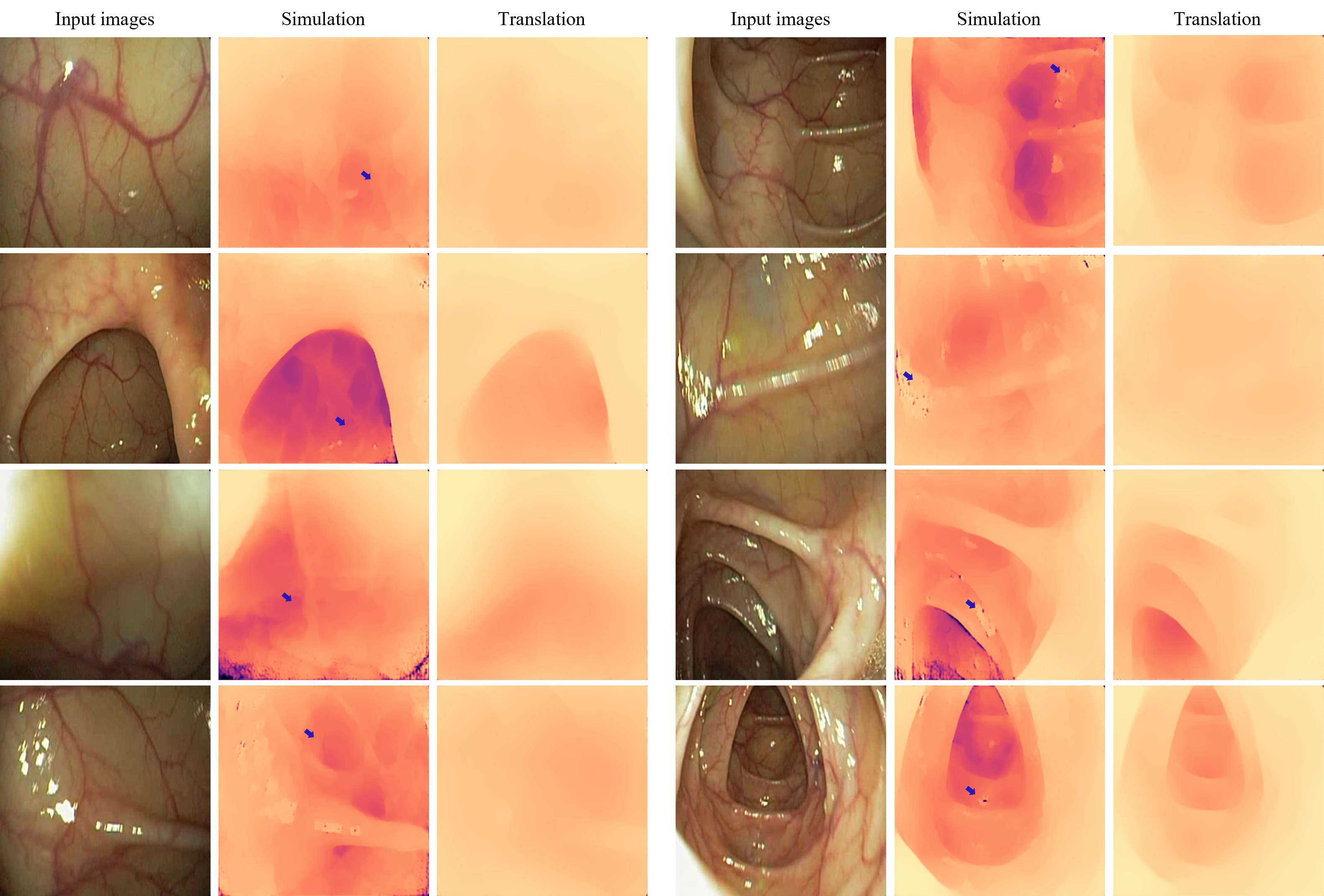}
\caption{The prediction results of the model trained with the simulation dataset and the model trained with the translated dataset. Blue arrows indicate texture errors in the simulation model.}
\label{fig:long}
\label{fig:onecol}
\end{figure}

The right side of figure.8 showed the depth inference error for the reflection of light. Images trained with simulation dataset appear as errors at the reflection point because they did not learn light reflection in the form of diffuse reflection by the surface. However, since the translated dataset generated diffuse type of light reflection, it is recognized as noise and no depth error appears.

\begin{figure}[htb!]
\includegraphics[width=\linewidth]{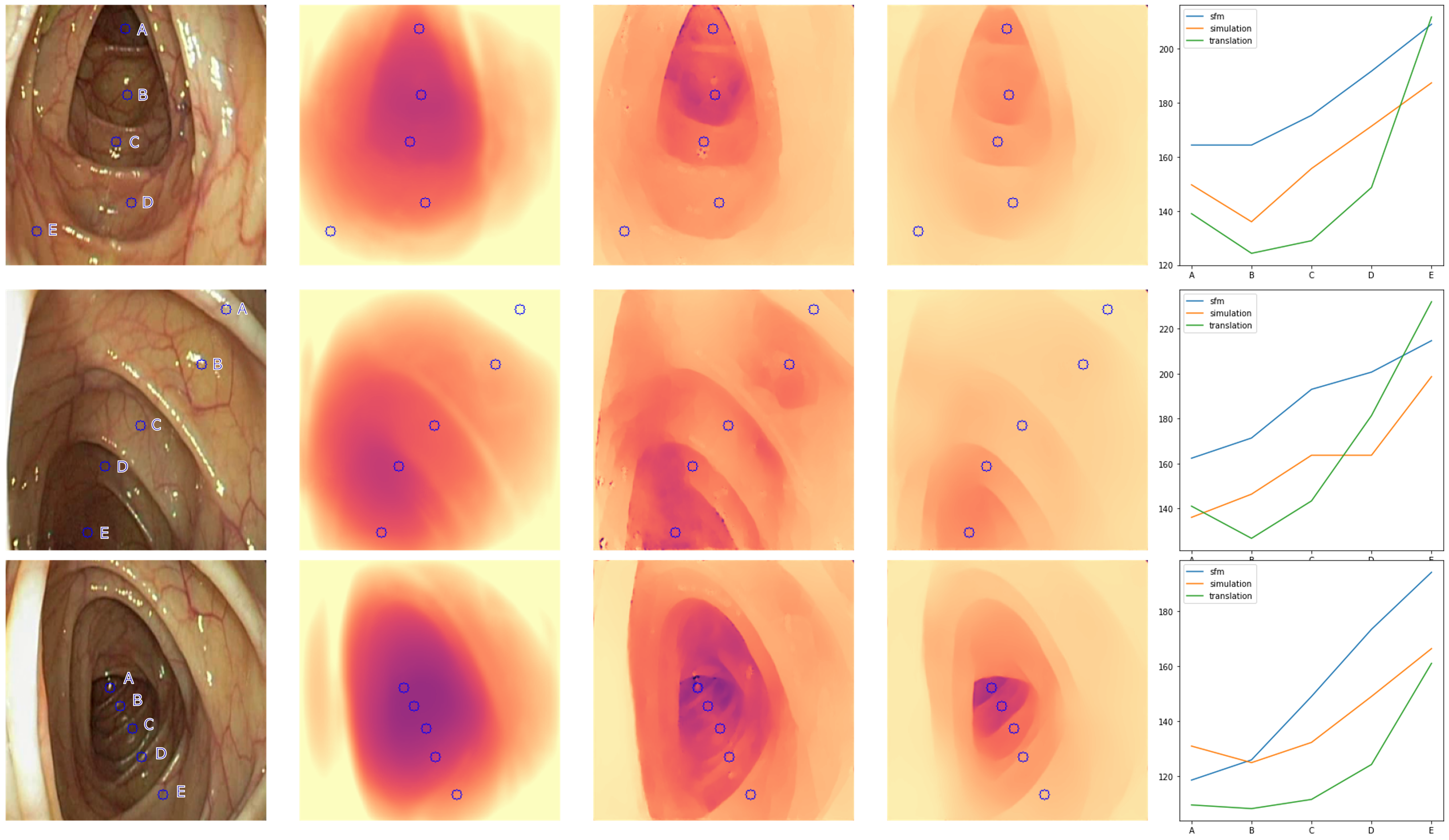}
\caption{Comparison of the heuristic depth with the predicted depth of each models.}
\label{fig:long}
\label{fig:onecol}
\end{figure}

Figure.9 represents selecting points in the direction in which the depth gradually decreases in human view from the real endoscopy to check whether a linear graph is generated when the corresponding points are listed in the depth results. The points of A to E picked in the endoscopy image are heuristic five points with depth gradually decreasing. The values of the same point were extracted from each depth inference result and expressed in a graph. The results of training with the translated dataset show linear results as expected. The graph increases because the depth is inverted to express it visually well. The results of the model trained with SfM or simulation dataset show steps or bowl shapes in the graph.

\section{Conclusion}
In this study, we proposed a quantitative and adaptive depth estimation for endoscopy using supervised learning methods with simulation environment and sim-to-real transfer deep learning. To generate a simulation environment similar to a real-world, we segment the colon from CT and use it as a model, and generate depth and simulation data computed through unity. The generated simulation data uses a method to perform sim-to-real transfer with real endoscopy images via cycleGAN to generate real endoscopy texture similar datasets. This generated data represents realistic endoscopy data with depth information. The translated data is used for network training to estimate depth images from images via the residual Unet. The trained network can be applied to the real endoscopy to measure depth. These depth images can be used to create a three-dimensional anatomical map and contribute to constructing a navigation system.

\section{Acknowledgment}

This research was supported by a grant of the Korea Health Technology R\&D Project through the Korea Health Industry Development Institute (KHIDI), funded by the Ministry of Health \& Welfare, Republic of Korea (grant number: HI19C0656030021).

\end{document}